\documentclass[notitlepage]{revtex4-1}
\usepackage[colorlinks=true,breaklinks=false]{hyperref}

\usepackage{graphicx}
\usepackage[normalem]{ulem}
\usepackage{amsmath,amstext,amssymb}
\usepackage{dcolumn}
\usepackage{bm}
\usepackage[english]{babel}
\usepackage{xcolor}

\newcommand{\FT}{{\mathcal F}}
\newcommand{\IFT}{{\mathcal F^{-1}}}
\newcommand{\Ea}{{\mathcal E}}
\newcommand{\Pa}{{\mathcal P}}

\begin{document}

\title{The Nonlinear Analytical Envelope Equation in quadratic nonlinear crystals}

\author{Morten Bache$^{1,*}$}

\affiliation{
$^1$DTU Fotonik, 
Technical University of Denmark,
  Bld. 343, DK-2800 Kgs. Lyngby, Denmark
$^*$Corresponding author: moba@fotonik.dtu.dk, \today }

\begin{abstract}
\noindent

\end{abstract}


\maketitle

We here derive the so-called Nonlinear Analytical Envelope Equation (NAEE) inspired by the work of Conforti et al. \cite{Conforti2013}, whose notation we follow. We present a complete model that includes $\chi^{(2)}$ terms (see \cite{conforti:2010PRA}), $\chi^{(3)}$ terms (see \cite{Conforti2013}), and then extend the model to delayed Raman effects in the $\chi^{(3)}$ term. We therefore get a complete model for ultrafast pulse propagation in quadratic nonlinear crystals similar to the Nonlinear Wave Equation in Frequency domain \cite{Guo:2013}, but where the envelope is modelled rather than the electrical field while still keeping a sub-carrier level resolution. The advantage of the envelope formation is that the physical origin of the additional terms that are included to model the physics at the carrier level becomes more clear, in contrast to the electric field equations that are more "black box" expansions of the electrical field. We also point out that by comparing our results to a very similar model and widely used model \cite{Genty2007}, the Raman terms presented there will most likely lead to an artificially lower Raman effect.

The basic forward propagation equation is
\begin{align}\label{eq:UPPE}
i\frac{\partial E_\omega}{\partial z}+\beta(\omega)E_\omega =
-\frac{\omega}{2cn(\omega)}P_{\rm NL}(\omega)
\end{align}
where
\begin{align}\label{eq:FT}
E_\omega(z)=\FT\{ E(t,z) \}\equiv\int_{-\infty}^\infty dt e^{+i\omega t}E(t,z)
\end{align}
The connection to the analytical (and complex) field $\Ea$ is
\begin{align}\label{eq:E-analytical}
\Ea(t,z)=\pi^{-1}\int_{0}^\infty d\omega e^{-i\omega t}E_\omega(z)
\end{align}
so the Fourier transform of the (real) electrical field is
\begin{align}\label{eq:FT-elec}
E_\omega(z)=(\Ea_\omega(z)+\Ea^*_\omega(z))/2
\end{align}
which is real. The analytical envelope satisfies
\begin{align}\label{eq:Eana}
\Ea_{\omega>0}=2E_\omega
\end{align}
as well as $\Ea_{\omega<0}=0$ and $\Ea_{\omega=0}=E_{\omega=0}$.

We will in what follows study only the case of a single pump field, so we drop the traditional general subscripts on all fields for the nonlinear parts. We also adopt the assumption of an isotropic nonlinear response and used scalar notation for simplicity. The nonlinear polarization is
\begin{align}\label{eq:P_NL}
P_{\rm NL,\omega}\equiv P_{\rm NL,\omega}^{(2)}+P_{\rm NL,\omega}^{(3)}
\end{align}
and in time domain they are given by \cite{hellwarth:1977}
\begin{align}\label{eq:P_NL-time-chi2}
P_{\rm NL}^{(2)}(t)&=\int_{-\infty}^\infty dt_1 \int_{-\infty}^\infty dt_2
\chi^{(2)}(t,t_1,t_2)E(t_1)E(t_2)
\\
\label{eq:P_NL-time-chi3}
P_{\rm NL}^{(3)}(t)&=\int_{-\infty}^\infty dt_1
\int_{-\infty}^\infty dt_2 \int_{-\infty}^\infty dt_3
\chi^{(3)}(t,t_1,t_2,t_3)E(t_1)E(t_2)E(t_3)
\end{align}
It is generally accepted that quadratic nonlinearities are instantaneous $\chi^{(2)}(t,t_1,t_2)=\chi^{(2)}\delta(t-t_1)\delta(t_1-t_2)$, so that $P_{\rm NL,t}^{(2)}=\chi^{(2)}E^2(t)$, meaning that in frequency domain  $P_{\rm NL}^{(2)}(\omega)=\chi^{(2)}\FT\{E^2(t)\}$. For the cubic case, the standard Born-Oppenheimer approximation implies that for optical fields one can write \cite[Eq. (4.9)]{hellwarth:1977}
\begin{align}\label{eq:P_NL-time-chi3-BO}
P_{\rm NL}^{(3)}(t)&=\chi^{(3)}_{\rm el}E^3(t)+
E(t)\int_{-\infty}^\infty ds \chi_R^{(3)}(t-s)E^2(s)
\end{align}
which is equivalent to taking $\chi^{(3)}(t,t_1,t_2,t_3)= \chi^{(3)}_{\rm el}\delta(t-t_1)\delta(t_1-t_2)\delta(t_2-t_3)
+\delta(t-t_1)\chi_R^{(3)}(t_1-t_2)\delta(t_2-t_3)$ \cite[Eq. (4.12)]{hellwarth:1977}. One can also conveniently define \cite[Eq. (2)]{blow:1989}
\begin{align}\label{eq:P_NL-time-chi3-BO-Blow}
P_{\rm NL}^{(3)}(t)&=\chi^{(3)}_{\rm tot}E(t)\int_{-\infty}^\infty ds g(t-s)E^2(s)
\\
g(t)&= \alpha\delta(t)+(1-\alpha) h_R(t)
\end{align}
where $g(t)$ is a normalized response function so that $\int_{-\infty}^\infty dt g(t)=1$ (which implies that $h_R(t)$ is normalized as well), and therefore $1-\alpha$ is the fractional nonlinear response of the Raman effect to the total nonlinearity $\chi^{(3)}_{\rm tot}$.

In frequency domain this can be written as
\begin{align}\label{eq:PNL-chi3-Raman-omega}
P_{\rm NL}^{(3)}(\omega)&=\chi_{\rm el}^{(3)}\FT\{E^3(t)\}+
\FT\left\{
E(t)\int_{-\infty}^\infty ds \chi_R^{(3)}(t-s)E^2(s)\right\}
\\
&=\chi_{\rm el}^{(3)}\FT\{E^3(t)\}+
\FT\left\{
E(t)\IFT[\tilde\chi_R^{(3)}(\omega-\omega_0)\FT\{E^2(t)\}]\right\}
\end{align}
where
\begin{align}\label{eq:chi3-R-omega}
\tilde\chi_R^{(3)}(\omega-\omega_0)=\FT\{\chi_R^{(3)}(t)\}
\end{align}
We note that the Raman response does not depend on the absolute laser frequency but on the frequency offset from the laser frequency $\omega-\omega_0$.


The envelope is now introduced as
\begin{align}\label{eq:envelope}
A(z,t)=\Ea(z,t)e^{-i\beta_0 z+i\omega_0t}
\end{align}
which means we can now evaluate the nonlinear polarization response in time domain by carrying out the expansions. Let us start with the instantaneous part
\begin{align}
P_{\rm NL,el}^{(3)}(z,t)&=\chi_{\rm el}^{(3)}\left[(\Ea(z,t)+\Ea^*(z,t))/2\right]^3
\nonumber\\
&=\chi^{(3)}_{\rm el}\left[(A(z,t)e^{-i\omega_0 t+i\beta_0z}+A^*(z,t)e^{+i\omega_0 t-i\beta_0z})/2\right]^3
\nonumber\\
&=\frac{3\chi^{(3)}_{\rm el}}{8}\left[
|A|^2Ae^{-i\omega_0 t+i\beta_0z}+
\tfrac{1}{3}A^3e^{-3i\omega_0 t+3i\beta_0z}+
\tfrac{1}{3}A^{*3}e^{3i\omega_0 t-3i\beta_0z}+
|A|^2A^*e^{i\omega_0 t-i\beta_0z}
\right]
\label{eq:PNL3-el}
\end{align}

By defining the analytical nonlinear polarization in the same way as for the field, i.e. $\Pa_{\rm NL,\omega>0}=2P_{\rm NL,\omega}$, we can introduce a nonlinear polarization envelope
\begin{align}\label{eq:envelope-P}
A_{\rm NL}(z,t)=\Pa(z,t)e^{-i\beta_0 z+i\omega_0t}
\end{align}
Using this, the instantaneous polarization envelope is
\begin{align}
A_{\rm NL,el}^{(3)}(z,t)=\frac{3\chi_{\rm el}^{(3)}}{4}\left[
|A|^2A+
\tfrac{1}{3}A^3e^{-2i\omega_0 t+2i\beta_0z}+
|A|^2A^*e^{2i\omega_0 t-2i\beta_0z}
\right]_+
\label{eq:ANL3-el-env}
\end{align}
where the '$+$' subscript indicates that only the positive frequencies must be accounted for. This is also why the $A^{*3}$ term is no longer present.

The Raman part is a bit more involved
\begin{align}
P_{\rm NL,R}^{(3)}(z,t)=&\frac{1}{8}(\Ea(z,t)+\Ea^*(z,t))
\left[\chi_R^{(3)}(t)*(\Ea(z,t)+\Ea^*(z,t))^2\right]
\nonumber\\
=&\frac{1}{8}(A(z,t)e^{-i\omega_0 t+i\beta_0z}+A^*(z,t)e^{+i\omega_0 t-i\beta_0z})
\left[\chi_R^{(3)}(t)* \left\{A(z,t)e^{-i\omega_0 t+i\beta_0z}+A^*(z,t)e^{+i\omega_0 t-i\beta_0z}\right\}^2\right]
\nonumber\\
=&\frac{1}{8}
(A(z,t)e^{-i\omega_0 t+i\beta_0z}+A^*(z,t)e^{+i\omega_0 t-i\beta_0z})
\nonumber\\&\times
\Big(
[\chi_R^{(3)}(t)*e^{-i2\omega_0 t+i2\beta_0z} A^2(z,t)]+
[\chi_R^{(3)}(t)*e^{i2\omega_0 t-i2\beta_0z} A^{*2}(z,t)]
+2[\chi_R^{(3)}(t) *|A(z,t)|^{2}]
\Big)
\label{eq:PNL3-R}
\end{align}
where "$*$" denotes convolution. For the analytical nonlinear polarization the term
\begin{align*}
A^*(z,t)e^{+i\omega_0 t-i\beta_0z}[\chi_R^{(3)}(t)* e^{i2\omega_0 t-i2\beta_0z} A^{*2}(z,t)]
\end{align*}
operates only at negative frequencies and is therefore neglected when writing the Raman polarization envelope
\begin{align}
A_{\rm NL,R}^{(3)}(z,t)&=\frac{1}{4}
\Big\{\left(A(z,t)+A^*(z,t)e^{+i2\omega_0t-i2\beta_0 z}\right)
[\chi_R^{(3)}(t)* \left(e^{-i2\omega_0 t+i2\beta_0z} A^2(z,t)+2|A(z,t)|^2\right)]
\nonumber\\
+&A(z,t)[\chi_R^{(3)}(t)*e^{i2\omega_0 t-i2\beta_0z} A^{*2}(z,t)]\Big\}_+
\label{eq:ANL3-R-env}
\end{align}

The Raman response one would get from a narrow-band sub-ps to ps pulse can be calculated if we keep only the terms that are only oscillating at the carrier frequency (i.e. where the polarization envelope does not have any $e^{\pm i\omega_0t}$ terms) \cite[equations below Eq. (8)]{blow:1989}
\begin{align}\label{eq:SVEA}
A_{\rm NL,el}^{(3)}(z,t)+A_{\rm NL,R}^{(3)}(z,t)&\simeq
\frac{3}{4}\chi_{\rm tot}^{(3)}A(z,t)\int_{-\infty}^\infty ds f(t-s)|A(z,s)|^2\\
f(t)&\equiv \alpha \delta(t)+\tfrac{2}{3}(1-\alpha) h_R(t)
\label{eq:R-alpha}
\end{align}
The material cubic nonlinearity is typically measured with exactly such pulses. Under the approximation that the pulse varies slowly compared to the Raman response function $h_R(t)$  we would get a polarization term $A_{\rm NL,el}^{(3)}(z,t)+A_{\rm NL,R}^{(3)}(z,t)\simeq \tfrac{3}{4}\chi_{\rm tot}^{(3)}|A|^2A [\alpha+\tfrac{2}{3}(1-\alpha)\int_{-\infty}^\infty dt' h_R(t-t')]$. Since $\int_{-\infty}^\infty dt' h_R(t-t')=\int_{-\infty}^\infty dt' h_R(t')=1$ we get $A_{\rm NL,el}^{(3)}(z,t)+A_{\rm NL,R}^{(3)}(z,t)\simeq \tfrac{3}{4}\chi_{\rm tot}^{(3)}|A|^2A$. Here we have defined the total cubic nonlinear susceptibility that one would measure e.g. using a z-scan technique where the SPM-induced nonlinear refraction is deduced $n_2\propto \chi_{\rm tot}^{(3)}$.
\begin{align}\label{eq:chi3-tot}
\chi_{\rm tot}^{(3)}=\chi_{\rm el}^{(3)}+\tfrac{2}{3}\tilde \chi_{\rm R}^{(3)}(0)
\end{align}
where $\tilde \chi_{\rm R}^{(3)}(0)=\int_{-\infty}^\infty dt' \chi_R^{(3)}(t')$ is the DC component of the frequency domain response function, which conveniently gauges the integral over the temporal response. Now we define
\begin{align}\label{eq:f_R}
f_R=\tfrac{2}{3}(1-\alpha)
\end{align}
to get the classical result \cite[Eq. (2.3.33)]{agrawal:2012}
\begin{align}\label{eq:SVEA-1}
A_{\rm NL,el}^{(3)}(z,t)+A_{\rm NL,R}^{(3)}(z,t)&\simeq
\frac{3\chi_{\rm tot}^{(3)}}{4}A(z,t)\int_{-\infty}^\infty ds R(t-s)|A(z,s)|^2\\
R(t)&\equiv (1-f_R)\delta(t)+f_Rh_R(t)
\end{align}
Thus the $f_R$ value is somewhat smaller than the $1-\alpha$ value; the latter is in \cite{blow:1989} reported to be 0.3 for silica, while typically $f_R=0.18$ is used for silica, in good agreement with the $1-\alpha=0.3$ value.

With these definitions the Raman polarization envelope Eq. (\ref{eq:ANL3-R-env}) becomes
\begin{multline}
A_{\rm NL,R}^{(3)}(z,t)=\frac{3\chi_{\rm tot}^{(3)}}{4} f_R
\Big\{
\tfrac{1}{2}A(z,t)\left[h_R(t)*e^{i2\omega_0 t-i2\beta_0z} A^{*2}(z,t)\right]
\\
+\left(A(z,t)+A^*(z,t)e^{+i2\omega_0t-i2\beta_0 z}\right)
\left[h_R(t)* \left(e^{-i2\omega_0 t+i2\beta_0z} \tfrac{1}{2}A^2(z,t)+|A(z,t)|^2\right)\right]
\Big\}_+
\label{eq:ANL3-R-env-hR}
\end{multline}
As a sanity check we now see that when discarding the oscillatory terms $\propto e^{\pm i2\omega_0 t}$ then we are left with the usual Raman response $A_{\rm NL,R}^{(3)}(z,t)=\frac{3\chi_{\rm tot}^{(3)}}{4} f_R \int_{-\infty}^\infty dt' h_R(t-t')|A(z,t')|^2$.

Finally we refer to \cite{conforti:2010PRA} for the polarization envelope in presence of a $\chi^{(2)}$ nonlinearity. Note the equations there have a different definition of the exponential sign of the forward Fourier transform and the envelope, as well as a different definition of the nonlinear polarization. Rewriting to our definitions:
\begin{align}
A_{\rm NL}^{(2)}(z,t)=\frac{\chi^{(2)}}{2}\left[
A^2e^{-i\omega_0 t+i\beta_0z}+
2|A|^2e^{i\omega_0 t-i\beta_0z}
\right]_+
\label{eq:ANL2-env}
\end{align}

Combining this into the forward propagation equation in time domain we get
\begin{multline}\label{eq:NLSE-NAEE}
i\frac{\partial A}{\partial \zeta}+\hat D_\tau A
+\frac{\chi^{(2)}\omega_0}{2n(\omega_0)c}\hat S_\tau \left[
\tfrac{1}{2}A^2 e^{-i\omega_0 \tau-i\Delta k_{\rm pg}\zeta}
+|A|^2 e^{i\omega_0 \tau+i\Delta k_{\rm pg}\zeta}
\right]_+
\\
+\frac{3\omega_0\chi^{(3)}}{8n(\omega_0)c}\hat S_\tau \Big[
(1-f_R)\left(|A|^2A+|A|^2A^* e^{i2\omega_0\tau + i2\Delta k_{\rm pg} \zeta}+ \tfrac{1}{3}A^3 e^{-i2\omega_0\tau - i2\Delta k_{\rm pg} \zeta}\right)
\\
+f_R \Big\{
\tfrac{1}{2}A(\zeta,\tau)
\int_{-\infty}^\infty d\tau'
h_R(\tau-\tau') e^{i2\omega_0 \tau'+i2\Delta k_{\rm pg}\zeta} A^{*2}(\zeta,\tau')
\\
+\left(A(\zeta,\tau)+A^*(\zeta,\tau)e^{i2\omega_0 \tau+i2\Delta k_{\rm pg}\zeta}\right)
\int_{-\infty}^\infty d\tau'
h_R(\tau-\tau')\left(\tfrac{1}{2}A^2(\zeta,\tau')e^{-i2\omega_0 \tau'-i2\Delta k_{\rm pg}\zeta}+|A(\zeta,\tau')|^2\right)
\Big\}
\Big]_+=0
\end{multline}
where for notational reasons we have suppressed the dependence of $A$ on $\zeta$ and $\tau$ except in the Raman part where it is spelled out for clarity. We have also switched to a moving reference frame $\zeta=z$ and $\tau=t-z\beta_1$. This gives rise to the NAEE phase-mismatch factor $\Delta k_{\rm pg}=\beta_1\omega_0-\beta_0$. In fact $\Delta k_{\rm pg}\zeta$ is the phase accumulated on the envelope due to the velocity mismatch between the carrier phase velocity and the envelope group velocity. To see this we use $\beta_1=1/v_{\rm g}=n_{\rm g}/c$ and $\beta_0=n_{\rm p}\omega_0/c$, where the "phase index" $n_{\rm p}=n(\omega_0)$, to write $\Delta k_{\rm pg}=
(n_{\rm g}-n_{\rm p})\omega_0/c=\omega_0(1/v_{\rm g}-1/v_{\rm p})$. Finally, $\hat D_\tau=\sum_{m=2}^\infty \beta_m(\omega_0)\left(i\tfrac{\partial}{\partial \tau}\right)^m/m!$ is the usual dispersion operator expansion and $\hat S_\tau=1+i\omega_0^{-1}\tfrac{\partial}{\partial \tau} $ is the self-steepening operator.

If we now introduce the auxiliary field $a(\zeta,\tau)=A(\zeta,\tau)e^{-i\Delta k_{\rm pg}\zeta}$ then the equation simplifies to
\begin{multline}\label{eq:NLSE-NAEE-DK}
i\frac{\partial a}{\partial \zeta}+(\hat D_\tau -\Delta k_{\rm pg})a
+\frac{\chi^{(2)}\omega_0}{2n(\omega_0)c}\hat S_\tau \left[
\tfrac{1}{2}a^2 e^{-i\omega_0 \tau}
+|a|^2 e^{i\omega_0 \tau}
\right]_+
+\frac{3\omega_0\chi^{(3)}}{8n(\omega_0)c}\hat S_\tau \Big[
(1-f_R)
\left(|a|^2a+|a|^2a^* e^{i2\omega_0\tau}+ \tfrac{1}{3}a^3 e^{-i2\omega_0\tau} \right)
\\
+f_R \Big\{
\tfrac{1}{2}a(\zeta,\tau)
\int_{-\infty}^\infty d\tau'
h_R(\tau-\tau')e^{i2\omega_0 \tau'} a^{*2}(\zeta,\tau')
\\
+\left(a(\zeta,\tau)+a^*(\zeta,\tau)e^{i2\omega_0\tau }\right)
\int_{-\infty}^\infty d\tau'
h_R(\tau-\tau')(\tfrac{1}{2}a^2(\zeta,\tau')e^{-i2\omega_0 \tau'}+|a(\zeta,\tau')|^2)
\Big\}
\Big]_+=0
\end{multline}
This is advantageous to model numerically because the cumbersome update of the phase mismatch in the nonlinear step is moved to the linear dispersion operator.

Finally, using the shorthand notation $\bar a(\zeta,\tau)=a(\zeta,\tau)e^{-i\omega_0 \tau}$ then the NAEE can be written on a more compact form
\begin{multline}\label{eq:NLSE-NAEE-DK-shorthand}
i\frac{\partial a}{\partial \zeta}+(\hat D_\tau-\Delta k_{\rm pg}) a
+\frac{d_{\rm eff}\omega_0}{n(\omega_0)c}\hat S_\tau \left[\tfrac{1}{2}a\bar a +a\bar a^* \right]_+
+\frac{3\omega_0\chi^{(3)}}{8n(\omega_0)c}\hat S_\tau \Big[
(1-f_R)a(|\bar a|^2+\bar a^{*2}+ \tfrac{1}{3}\bar a^2 )
\\
+f_R \Big\{\tfrac{1}{2}a(\zeta,\tau)
\int d\tau'
h_R(\tau-\tau') \bar a^{*2}(\zeta,\tau')
+e^{i\omega_0\tau}2{\rm Re}\left(\bar a(\zeta,\tau)\right)
\int d\tau'
h_R(\tau-\tau')(\tfrac{1}{2}\bar a^2(\zeta,\tau')+|\bar a(\zeta,\tau')|^2)
\Big\}
\Big]_+=0
\end{multline}
where as per usual $d_{\rm eff}=\chi^{(2)}/2$. This notation is useful also for practical reasons as numerically the vector $e^{i\omega_0\tau}$ is applied only twice in the nonlinear step, namely for calculating $\bar a$ and in front of the last Raman term.

A similar equation for the Raman terms was reported in \cite{Genty2007}. The equation they use have the nonlinear cubic terms written as
\begin{align}
&(1-f_R)\left(|U|^2U+\tfrac{1}{3}U^3 e^{-i2\omega_0t}\right)
\nonumber\\
&+\tfrac{2}{3}f_R \Big\{
U(z,t)
\int_{-\infty}^\infty dt'
h_R(t-t') |U(z,t)|^2
+e^{-i\omega_0 t}{\rm Re}\left(U(z,t)e^{-i\omega_0 t}\right)
\int_{-\infty}^\infty dt'
h_R'(t-t')U^2(z,t')
\Big\}
\label{eq:NLSE-Genty}
\\
=&(1-f_R)\left(|U|^2U+\tfrac{1}{3}U^3 e^{-i2\omega_0t}\right)
\nonumber\\
&+\tfrac{2}{3}f_R \Big\{
U(z,t)
\int_{-\infty}^\infty dt'
h_R(t-t') |U(z,t)|^2
+e^{i\omega_0 t}{\rm Re}\left(U(z,t)e^{-i\omega_0 t}\right)
\int_{-\infty}^\infty dt'
h_R(t-t')e^{-i2\omega_0 t'}U^2(z,t')
\Big\}
\label{eq:NLSE-Genty1}
\end{align}
where they define $h_R'(t)=h_R(t)e^{i2\omega_0t}$. It is important to note that this equation is reported in the stationary lab frame and that the envelope ansatz is $E_+=\tfrac{1}{2}[A_+(z,t)e^{i\omega_0t}+A_+^*(z,t)e^{-i\omega_0t}]$, i.e. it does not include the $e^{-i\beta_0z}$ term. Moreover, $U$ is normalized so $|U|^2$ gives the power. They also neglect the terms related to $UU^{*2}$ both in the SPM and Raman part.
Despite this, we see some similarities: The SPM and THG terms seem the same, and components of the Raman part are recognizable: both the standard term in Raman related to $|U|^2$ and the non-standard Raman term related to $U^2$ have the same forms as in our case. That being said, the factor $\tfrac{2}{3}f_R$ in front of the Raman terms seems to indicate that the "classical" limit of Eq. (\ref{eq:SVEA-1}) has not been imposed, and that the value $f_R$ they report is identical to the $1-\alpha$ value used by \cite{blow:1989}, see Eq. (\ref{eq:R-alpha}). This means that the Raman effect is underestimated by a factor of $2/3$ when using classical values of the Raman effect, such as the value $f_R=0.18$ for silica that they use, leading to an artificial, smaller Raman effect.

\bibliographystyle{apsrev4-1}
\bibliography{m:/homeextra/Bibtex/literature}

\end{document}